\documentclass[prl,twocolumn]{revtex4}

\usepackage{graphicx}
\usepackage{dcolumn}
\usepackage{amsmath}
\usepackage {epsfig}

\def\cf2{$\beta^{\prime}-(ET)_2SF_5CF_2SO_3$~}

\makeatother

\begin{document}
\title[Short Title]{High field magnetic resonant properties of \cf2}
\author{I.B. Rutel$^1$, S.A. Zvyagin$^1$, J.S. Brooks$^1$, J. Krzystek$^1$, P. Kuhns$^1$, A.P. Reyes$^1$, E. Jobiliong$^1$, B.H. Ward$^2$,J.A. Schlueter$^3$, R.W. Winter$^4$, and G.L. Gard$^4$}
\affiliation{$^1$National High Magnetic Field Laboratory and
Physics Department, Florida State University, Tallahassee, FL
32310} \affiliation{$^2$ Florida State University, Department of
Chemistry and Biochemistry, Tallahassee, FL
32306-4390}\affiliation{$^3$Materials Science Divisions, Argonne
National Laboratory, Argonne, IL 60439-4831}
\affiliation{$^4$Department of Chemistry, Portland State
University, Portland, OR 97207-0751}
\date{\today}

\begin{abstract}
The charge transfer salt \cf2, which has previously been
considered a spin-Peierls material with a $T_{SP}\sim$33 K, is
examined using high-resolution high-field
sub-millimeter/millimeter wave electron spin resonance (ESR), and
nuclear magnetic resonance (NMR) techniques. A peak in the nuclear
spin-lattice relaxation behavior in fields of 8 T, accompanied by
a broadening and paramagnetic shift of the resonance line,
indicates a phase transition at $T_c\sim$20 K. A pronounced change
in the high-field ESR excitation spectra at $\sim$24 T, observed
at $T_c\sim$20 K, may indicate the onset of antiferromagnetic
(AFM) correlations of the low temperature phase in \cf2.
Peculiarities of the low-temperature magnetic and resonance
properties of \cf2 are discussed.
\end{abstract}

\maketitle

\section{Introduction}
The $(ET)_2X$ charge transfer salts represent a large number of
materials that possess properties ranging from insulator to
superconductor ($ET$ is bis(ethylenedithio)-tetrathiofulvalene).
The same can be said for the sub-class of compounds
$(ET)_2SF_5RSO_3$, where $R=CH_2, CF_2, CHF,
CH_2CF_2$\cite{wardandschlueter}. Modifications of the anion
result in fundamental changes in the structural and physical
properties, and investigations on these materials have been
pursued to determine the relation of the anion to the material
behavior. Some interest is focused on the anion relation to the
conduction path formed by $\pi - \pi$ orbital overlap of sulfur in
the ET donor molecules\cite{mori}-\cite{mold}. The role of the
anion in electrical properties is still unclear. However, it is
certain that the anion plays a role in partially determining the
structure of the compound and accepting an electron from the ET
molecule, forming a potential conduction hole. The hole then
promotes the spin character of the system, providing magnetic
interaction through band magnetism. It is interesting to
investigate such systems that show a transition of a low
dimensional magnetism into a three dimensional (3D) AFM state, as
well as those which exhibit a transition into a spin-Peierls
state.

\begin{figure}[]
\epsfig{file=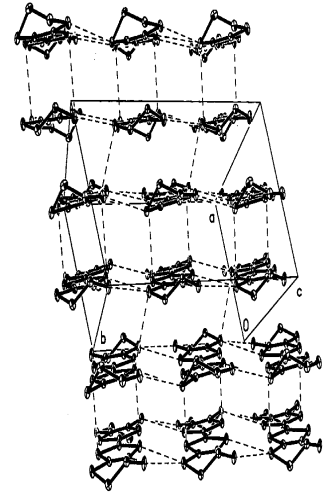,height=6.5cm} \caption{\cf2 cation
layer. The view is down the long axis of the molecule. Dashed
lines between donor stacks indicate $S \cdots S$ van der Waals
contacts less than 3.6 \AA. The ET molecules are dimerized and
form linear chains along the \emph{b}-axis\cite{wardandrutel}.
Interstack distances between donors (along \emph{b}-axis) are
$\sim$5.3 \AA, while intrastack distances are $\sim$7.3 \AA. The
\cf2 crystal cell has \emph{P}$\bar{1}$ symmetry below 120 K, and
forms a triclinic unit cell with \emph{a}= 7.699 \AA, \emph{b}=
13.151 \AA, \emph{c}= 17.643 \AA, and $\alpha $= 81.323$^{\circ}$,
$\beta $= 87.416$^{\circ}$, $\gamma $=
74.994$^{\circ}$\cite{wardandschlueter}.}\label{fig1}
\end{figure}
\cf2 is characterized by pairs of $(ET)^{+1/2}$ molecules which
contribute a single electron to the $SF_5CF_2SO_5^{-1}$ anion
acceptor, forming a charge transfer salt, which stacks with
face-to-face ET contacts (Fig.\ref{fig1}). The ET dimers form
linear spin chains along the \emph{b}-axis\cite{wardandrutel}.

\begin{figure}[]
\epsfig{file=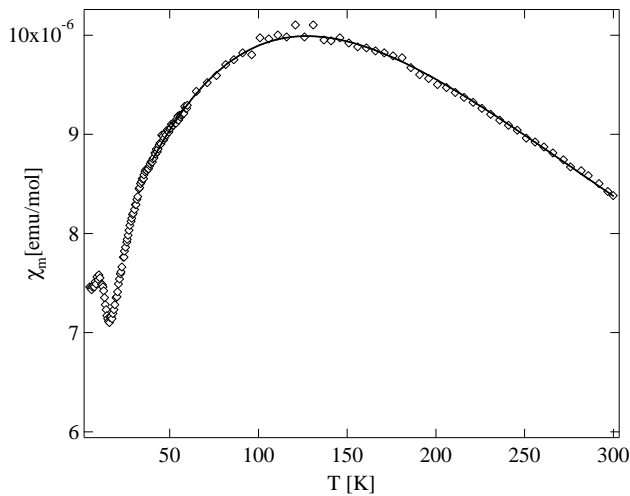,height=8.25cm, angle=-90} \caption{Magnetic
susceptibility versus temperature for \cf2. Solid curve is
Bonner-Fisher theory fit for S=$\frac{1}{2}$ HAF.}\label{fig1n}
\end{figure}

A measurement of the SQUID susceptibility for \cf2
(Fig.\ref{fig1n}) shows a broad maximum at T$\approx$115 K
indicating the low dimensional character of the interactions.
The susceptibility was fit using the Bonner-Fisher model for
S=$\frac{1}{2}$ Heisenberg antiferromagnet (HAF)\cite{bonner}. The
best fit for $T > 40$ K was obtained using a coupling constant
$J/k_B$ $=$ 352 K. Thus, the Bonner-Fisher fit together with a
crystal structure analysis suggests a $S=\frac{1}{2}$, linear
magnetic chain\cite{rutel}.

A kink in the susceptibility is observed at T$\sim$45 K (which was
suggested to be a spin-Peierls transition temperature
\cite{wardandschlueter}), followed by a pronounced drop at
T$\sim$20 K. As temperature drops below 20 K there is an intrinsic
feature which makes it difficult to reveal the nature of the
ground state in the low temperature regime.  Previous V-band ESR
experiments revealed a collapse of the gapless g$\sim$2 mode below
33 K. The change in the excitation spectrum was interpreted to as
possible evidence of the spin-Peierls
transition\cite{wardandrutel}. A change in the FIR properties was
observed below T$\sim$45 K, which indicate weak lattice
distortions in \cf2\cite{pigos}. In order to study the nature of
the low temperature transition in \cf2, we decided to perform ESR
experiments in an extended frequency-field range, using high-field
high-resolution submillimeter/millimeter wave ESR spectrometer
recently developed at the National High Magnetic Field Laboratory
(NHMFL), in Tallahassee, FL\cite{zvaygin}. In addition,
low-temperature properties of \cf2 are probed using a pulsed NMR
spectroscopy technique.

\section{Experimental Techniques}
The ESR method is a powerful tool in studying the nature of the
ground states in solids (see, for instance, refs. \cite{dumm1},
\cite{dumm2}, and \cite{sergei2}). We investigate the \cf2 system
employing ESR and NMR techniques at high frequencies and magnetic
fields. High resolution ESR was carried out in the 25 T W.M. Keck
resistive magnet (with magnetic field homogeneity, 12 ppm/cm DSV)
at the NHMFL, using backward wave oscillator (BWO) sources.
Investigations were performed by holding temperature and frequency
constant and sweeping the magnetic field. Details of the
high-field millimeter and submillimeter wave facility can be found
in Zvyagin, \emph{et al.}\cite{zvaygin}.

\begin{figure}[]
\epsfig{file=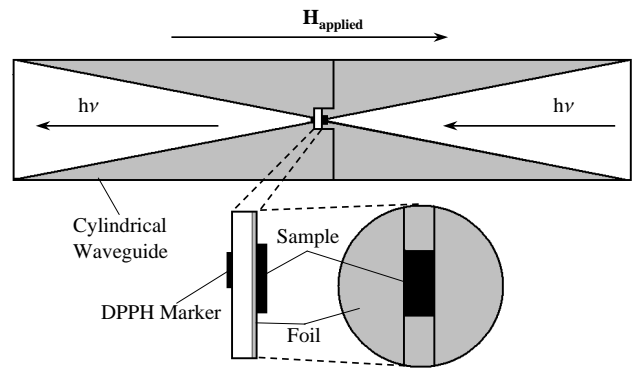,height=8.25cm,angle=-90} \caption{Direct
transmission ESR sample holder. The microwave radiation is
conically reduced to the sample size. The sample is placed on a
polymer holder with metallic masking to block any radiation not
passing through the sample. The marker, DPPH, is placed on the
side of the holder opposite from the sample.}\label{fig2}
\end{figure}

In our experiments we used single-crystalline samples with a
typical size of 0.5x0.1x0.1 $mm^3$\cite{wardandschlueter}. To
measure such small samples a special sample holder was built. It
consists of two metallic horns, conically reducing the microwave
radiation down to the sample size (Fig.\ref{fig2}). A metallic
foil was placed around the sample to reduce noise caused by
background radiation. The sample was placed in the Faraday
configuration, i.e. the propagation vector of the radiation
$\vec{k_{\omega}}$ is parallel to the external magnetic field
$\textbf{H}$. An optical chopper was used to modulate the
microwave power. $\textbf{H}$ was oriented perpendicular to the
\emph{c}-axis of the sample. The probe was coupled to the source
via circular waveguide sections and directed into the sample
space, where the microwave excitation passed through the sample, a
polymer platform, and then a marker compound of DPPH. The
radiation was then directed out of the probe and into a helium
cooled InSb detector.

NMR investigations were performed in a 15 T superconducting
magnet. A home-built MagRes2000 spectrometer\cite{reyes} and a
capacitively tuned and matched probe were employed. The
investigation focused on two nuclei, the $^{19}F$ nucleus on the
anion, and $^1H$ in the ethylene groups on the donor molecule(ET).
Remarkably, the NMR results revealed similar behavior for these
two sites which are also independent of field (or frequency),
suggesting a common mechanism for the spin dynamics at the anion
layer and at the ET molecule. For brevity, we present a
representative graph showing the results of the 340 MHz (8T)
investigation of the $^1$H nuclei.

\section{Results and Discussion}
We first present the NMR data performed in the temperature range
$\sim$3-300 K and show evidence indicating that the transition
seen at T$\sim$20 K  is not spin-Peierls in nature. The
characteristic behavior of the nuclear spin-lattice relaxation
rate ($T_1^{-1}$) in a prototypical spin-Peierls material
$CuGeO_3$ is described in Ref.\cite{fagot}. Above the transition
temperature $T_{sp}$, where the system would be in a uniform
phase, the fluctuations from weakly interacting spins contributes
to a large but weakly temperature-dependent relaxation rate.  In
$CuGeO_3$\cite{fagot}, this behavior persists until the
temperature drops below $T_{sp}$, where $T_1^{-1}$ decreases
sharply (by several orders of magnitude). There is also a change
in the magnetic hyperfine shift which follows the $T_1^{-1}$
curve, decreasing below the transition temperature.

\begin{figure}[]
\epsfig{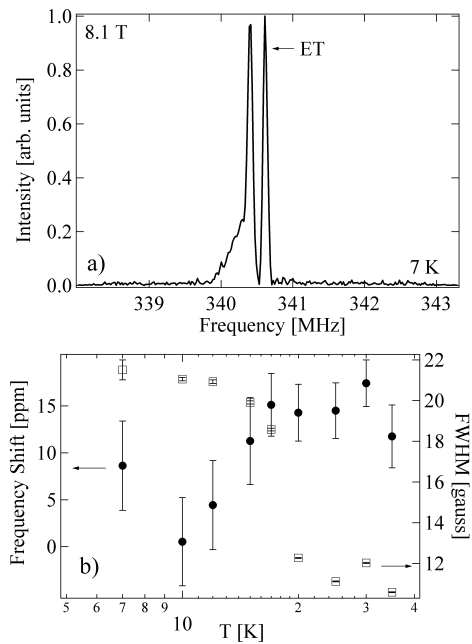}
\caption{a) NMR spectra for $^1$H at 340.61 MHz. The line marked
``ET" is the signal from the sample, the second line at lower
field is an off resonance spurious signal from the probe. b) The
hyperfine magnetic shift (left axis) for the \cf2 material,
determined from the spectral shift away from the carrier frequency
in the NMR investigation, and the full width at half maximum
(FWHM) analysis (right axis) of the spectral line for the title
compound from the NMR spectra. }\label{fig4}
\end{figure}

A typical NMR 1H spectrum in \cf2 is shown in Fig.\ref{fig4}a,
taken at 8.1 T. The data for $^{19}F$ NMR is not shown.  The $^1H$
data shows a narrow ET signal and a spurious signal at a lower
frequency. This second signal originates from the probe which was
difficult to eliminate completely. This signal has a much longer
$T_1$ than the ET signal and remains well resolved and separated
throughout the investigation.

\begin{figure}[]
\epsfig{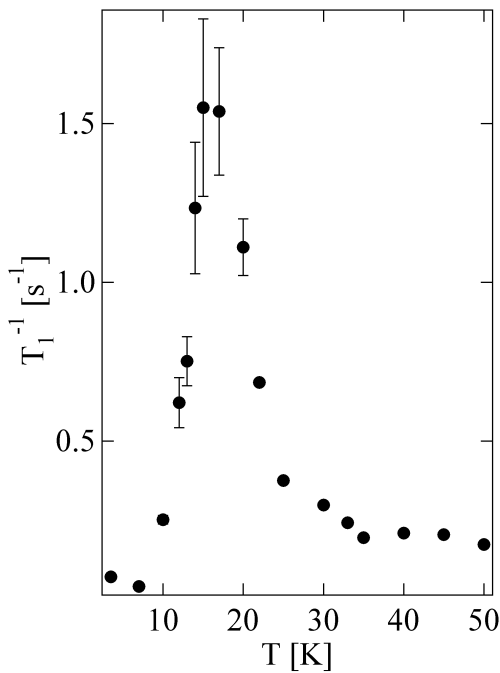} \caption{The
NMR data for the spin-lattice relaxation ($T_1^{-1}$) plotted
versus temperature for $^1H$ at 360 MHz. This investigation
clearly shows non spin-Peierls behavior. See text for
details.}\label{fig3}
\end{figure}

In Fig.\ref{fig3} we show the temperature dependence of $^1H$
$T_1^{-1}$.  At high temperatures, the relaxation rate is slightly
temperature-dependent. As the temperature is lowered, instead of
the expected drop at $T_{sp}$, we observed a sharp peak reaching a
maximum at around 20 K followed by a sudden drop at lower
temperatures.  Concomitantly, the FWHM, shown in Fig.\ref{fig4}b,
also increases around 20 K, while the resonance frequency
(Fig.\ref{fig4}b) drops. These data is suggestive of some form of
magnetic order. Indeed, this sample shows remarkably similar
behavior to the $LiMn_2O_4$ system, a material with an AFM ground
state.

The data for the $^{19}$F NMR (not shown) also reveals the same
peak in $T_1^{-1}$ at the same temperature. The similarity in
these behavior at two different sites, both the $^1H$ and $^{19}F$
(on the ET molecule and anion respectively) suggests that the
relaxation at both sites is dominated by a single spin
dynamics\cite{rutel}.

A representative transition to an AFM phase is given by Sugiyama
\emph{et al.} in the $LiMn_2O_4$ system\cite{sugi}.  In the case
of AFM ordering a peak is found in the $T_1^{-1}$ scattering with
the N\'{e}el temperature ($T_N$) indicated by the maximum of the
peak. Both the full-width at half maximum (FWHM) and the magnetic
hyperfine shift show increases at $T_N$.

Generally, a 3D antiferromagnetic ordering produces large
hyperfine fields at the nuclear site, thus shifting the resonance
and rendering the NMR line unobservable in the detected frequency
range.  In the present case, the signal remains tractable down to
the lowest temperature measured. This may occur if the ordering is
not long-ranged in nature or if the sample is not uniform, which
is unlikely.  It is also important to note that the bulk
susceptibility does not show any direct signature of AFM ordering.
In addition, although the FWHM follows the correct trend, it is
two orders of magnitude smaller than the $LiMn_2O_4$ data.
Nevertheless and whatever the case maybe, it is clear from the NMR
data the absence of any transition at $T_{sp}$ ($\sim$40 K) and
that the anomaly at $\sim$20 K is definitely something other than
a spin-Peierls transition.

We now present a discussion of the ESR investigation. In the case
of the low-frequency ESR (V-band and W-band) we have a single
resonance line which remains relatively constant in width,
amplitude and frequency above 20 K and shows a diminishing
amplitude, slight broadening, and small shift to lower field below
20 K. The resonance field data, H$\parallel$\emph{c}-axis also
show a relatively constant value with a slight change to lower
field below 20 K\cite{wardandrutel}. The experiment was repeated
using the sample investigated in both the NMR and high-frequency
ESR (see below). Figure \ref{figlowepr} reports data taken at 98
GHz, where similar shifts, broadening, and diminishing amplitude
are evidence for the reproducibility of the previous
investigation\cite{wardandrutel}.

\begin{figure}[]
\epsfig{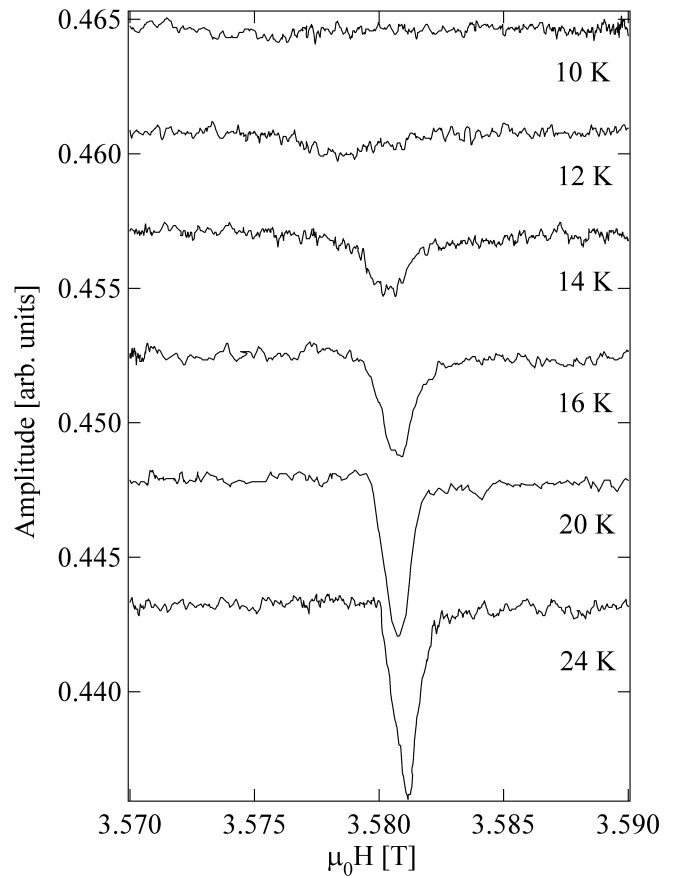}
\caption{Data from ESR performed at 98 GHz and 3.5 T.  The run was
performed on the same sample as the high-frequency investigation,
after the BWO experiment. The data shows reliable reproducibility
for the low-frequency behavior of the \cf2
sample.}\label{figlowepr}
\end{figure}

\begin{figure}[]
\epsfig{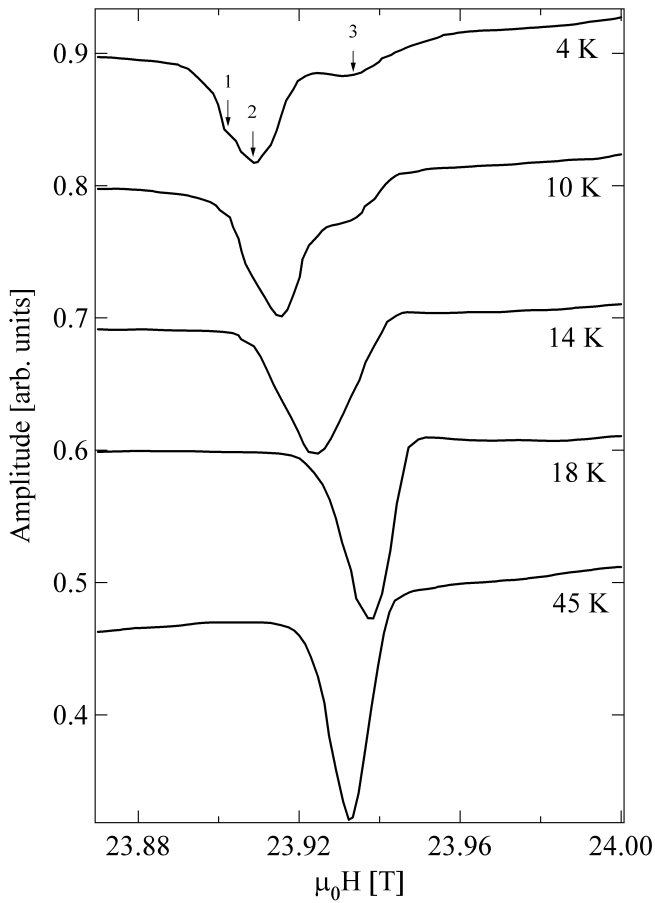}
\caption{Raw data from ESR investigation at 673 GHz for varying
temperature. The onset of broadening and introduction of new lines
with decreasing temperature suggests an AFM ordering observed
below the transition. Arrows indicate the multiple resonance lines
seen in the investigation. The spectra are offset for
clarity.}\label{fig5}
\end{figure}

This behavior is quite different from that observed at the higher
frequency and field. We first discuss the ESR results seen as the
raw data in Fig.\ref{fig5}, which shows the sample response with
changing temperature. At higher temperatures we observe one
signal, corresponding to the sample absorption. As the temperature
decreases we find the sample absorption line broadens and
eventually splits, where the onset of the broadening occurs around
20 K. Following the line to lower temperatures, the broadened line
begins to resolve into two, and then three distinct lines (denoted
by arrows 1, 2, and 3 in Fig.\ref{fig5}).

The high-frequency case (broadening followed by splitting) can be
explained by the appearance of three AFM modes at some fixed
frequency in the fields: below $H_{spin-flop}$, above
$H_{spin-flop}$ and in an intermediate critical regime at $H\sim
H_{spin-flop}$\cite{turov}, which were observed experimentally for
instance in $MnF_2$\cite{hagiwara}. The high-frequency-field data
indicate the appearance of new excitation modes, that may be
understood in terms of the model developed for spin excitations in
a uniaxial AFM with H$\parallel$\emph{easy}-axis\cite{turov}. Of
course, tilting  of the principal axis (even very small) and a low
symmetry (P1) of \cf2 can cause some deviations from the
model\cite{turov}.

\begin{figure}[]
\epsfig{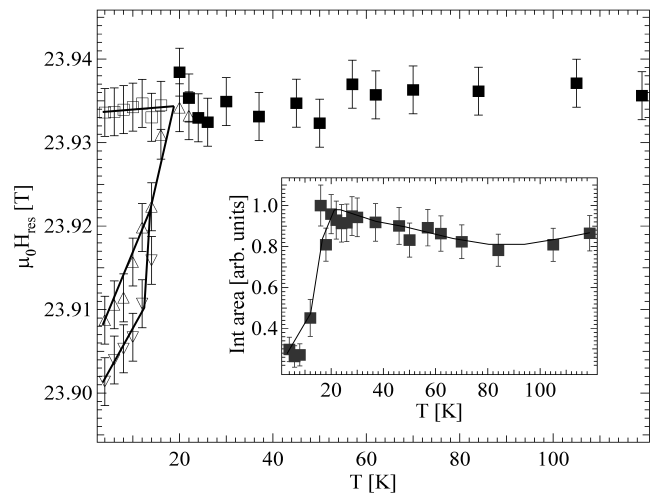} \caption{ESR resonance field versus temperature
for \cf2 oriented with $H\perp$\emph{c}-axis, the frequency is 673
GHz. Open triangles down and up, and open squares represent the
absorption lines denoted by arrows 1, 2, and 3 respectively, below
the transition temperature (Fig.\ref{fig5}). Solid squares
represent the absorption line in the paramagnetic phase, and solid
curves are guides for the eye. Inset: Integrated area of the ESR
absorption at the frequency 673 GHz plotted versus temperature.
The area is calculated from a multiple lorenzian fit for
absorption lines denoted by squares in Fig.\ref{fig5}. A drop in
the integrated intensity is seen below 20 K.}\label{fig6}
\end{figure}

Both sets of data show the onset of a transition at T$\sim$20K,
where the transition temperature is also verified in the
high-frequency data by the onset of splitting and remaining area
under the curve seen in Fig.\ref{fig6} and Fig.\ref{fig6}(inset).
However, several pronounced differences appear between the two
sets of data.  The first difference is the persistence of
excitations below about 10 K in the high-frequency data seen in
Fig.\ref{fig6} as opposed to that reported in
Ref.\cite{wardandrutel}.  Second, the high-frequency data show the
splitting of the line at low temperatures (Fig.\ref{fig6}), which
may indicate the onset of 3D AFM ordering.

Remarkably, in both cases the behavior is qualitatively similar to
that for AFM ordering.  The low-frequency case (broadening and
shifting) corresponds to similar features reported by Mitsudo
\emph{et al.} in $LaMnO_3$\cite{mitsudo}. The broadening is due to
1) the enhancement of the AFM correlations and 2) the change of
the frequency-field slope.    This leaves us with the conclusion
that both the low-frequency and the high-frequency results may
suggest an AFM type ordering in \cf2 at low temperature.

We should also make mention of the intrinsic nature of the anomaly
in the low-T portion of the susceptibility.  Assume that the low-T
part of the susceptibility is caused by the paramagnetic
impurities (chain ends, defects, other impurities).  This impurity
signal (which appears to be non-negligible in the susceptibility)
should manifest itself in the ESR as excitations with g$\sim$2. We
did see  the signal from such impurities for T$<$ 10 K, where the
98 GHz investigation and Ward \emph{et al.} did not.  This
suggests that the low-T portion is intrinsic to the system, and
may have something to do with the AFM ordering suggested by the
ESR.

Let us also discuss the possibility of an incommensurate phase. Is
is known that if we apply a high field we can enter into an
incommensurate phase, where the magnetic structure is
incommensurate with the crystal structure.  ESR was studied by W.
Palme \emph{et al.} in the incommensurate phase of the
spin-Peierls material $CuGeO_3$, where hysteresis appears to be a
basic property of the system\cite{palme}. However, in our
investigations we observed no hysteresis.  This would again
support the non-spin-Peierls nature of the $\sim$20 K transition.

\section{Conclusion}
We investigate \cf2, employing both pulsed NMR and high frequency
ESR methods. Both the NMR and ESR data provide evidence that the
title compound does not exhibit behavior characteristic for a
conventional spin-Peierls transition. We submit that both the NMR
and the high-frequency ESR investigations provide inconclusive
results as to the nature of the ground state in the low-T regime,
although there is some evidence of a 3D AFM ordering. It is clear
that there is a disparity between the low and high frequency/field
dependence, but at present no theory appears to describe the
behavior fully. Further investigations using x-ray diffraction and
neutron scattering could provide insight into a more exact
description of the crystallographic and magnetic structure through
the $T<20$ K phase transition.

\begin{acknowledgments}The FSU acknowledges support from NSF-DMR 99-71474 and IHRP
500/5031. A portion of this work was performed at the National
High Magnetic Field Laboratory, which is supported by NSF
Cooperative Agreement No.DMR-0084173 and by the State of Florida.
Work at Argonne National Laboratory was supported by the Office of
Basic Energy Science, Division of Materials Sciences, US
Department of Energy, under Grant W-31-109-ENG-38. Research as
Portland State University was supported by NSF Grant CHE-09904316.
\end{acknowledgments}

\end{document}